\documentclass[aps,showpacs,twocolumn]{revtex4-2}
\usepackage{amsmath}
\usepackage{amssymb}
\usepackage[dvipsnames]{xcolor}
\usepackage[many]{tcolorbox}
\usepackage{array}
\usepackage{multirow}
\usepackage{makecell}
\usepackage{hyperref}
\usepackage{booktabs}
\usepackage{array}

\hypersetup{hypertex=true, colorlinks=true, linkcolor=blue, urlcolor=blue, citecolor=blue}

\begin{document}

\title{Pair Filters in an Extended Hubbard Model at Resonance}
\author{E. S. Ma and Z. Song}
\email{songtc@nankai.edu.cn}
\affiliation{School of Physics, Nankai University, Tianjin 300071, China}
\begin{abstract}
We investigate the dynamics of bound pairs in an extended Hubbard model at
resonance. We show that a single fermion and a singlet bound pair possess
identical dispersion relations and scatter off each other. In contrast, a
neighboring doublon pair is dynamically pinned in the strong-interaction
regime. Remarkably, a single fermion and a singlet bound pair exhibit
fundamentally different scattering behaviors when encountering a pinned
doublon pair acting as a scattering center. While the singlet bound pair
undergoes perfect transmission, the single fermion is completely reflected.
These results demonstrate that a neighboring doublon pair functions as an
efficient filter that separates bound pairs from single particles. Numerical
simulations fully support the analytical predictions. Our findings provide a
dynamical mechanism for generating and manipulating bound-pair states.
\end{abstract}

\maketitle

\section{Introduction}

\label{Introduction}

The Hubbard model is one of the fundamental paradigms for understanding
strongly correlated quantum many-body systems. Despite its simple form, the
competition between kinetic energy and interactions gives rise to a rich
variety of emergent phenomena, including Mott insulating phases, magnetism,
unconventional superconductivity, and quantum phase transitions~Mott
insulators, magnetism, superconductivity and quantum phase transitions \cite%
{Hubbard1963,Essler2005,Bloch2008,Gross2017}. Owing to its relevance to
correlated electronic materials, ultracold atoms in optical lattices, and
programmable quantum simulators, the Hubbard model and its extensions
continue to occupy a central position in condensed-matter physics.

Among these extensions, the extended Hubbard model, which incorporates
nearest-neighbor interactions in addition to the on-site interaction,
exhibits substantially richer physics than the conventional Hubbard model.
Depending on the interaction strength and lattice geometry, nearest-neighbor
interactions can stabilize charge-density-wave phases, enhance
unconventional superconductivity, and induce various competing ordered
states~\cite%
{qu2022spin,Jiang2022enhancing,Peng2023enhanced,cao2025dominant,Kennedy2025extended}%
. More recently, the extended Hubbard model has also emerged as an
attractive platform for studying interaction-induced localization, kinetic
constraints, and Hilbert-space fragmentation, thereby revealing
unconventional nonequilibrium dynamics beyond the standard ergodic paradigm~%
\cite%
{Sala2020,Khemani2020,Frey2022,Moudgalya2022,He2026Hilbert,liu2026condensate}%
.

Besides equilibrium properties, the nonequilibrium dynamics of interacting
quantum systems has become a major research direction. In strongly
interacting lattices, particles may bind together to form stable composite
quasiparticles known as doublons. Owing to energy conservation, doublons can
remain remarkably long-lived in the strong-coupling regime, even though they
correspond to excited states of the system. Their existence was first
observed experimentally in ultracold atoms loaded into optical lattices~\cite%
{Winkler2006,Petrosyan2007,Valiente2008,Sensarma2010,Strohmaier2010,Preiss2015}%
, which stimulated extensive theoretical studies of repulsively bound pairs,
quantum walks, coherent transport, and collision dynamics of composite
particles~\cite%
{Petrosyan2007,Valiente2008,Jin2009coherent,jin2011fast,Jin2011perfect}.
Understanding how these composite objects scatter from other many-body
excitations is not only of fundamental interest but also provides promising
opportunities for coherent quantum-state manipulation.

Previous studies have mainly focused on scattering processes involving
static impurities, local defects, or external potentials. Comparatively less
attention has been devoted to interaction-induced scattering in which one
composite excitation itself acts as a dynamical scattering center for
another. Such processes are particularly interesting because the internal
structure of composite quasiparticles can fundamentally modify their
transport properties, leading to scattering behaviors that have no
counterpart in single-particle physics. Exploring these interaction-induced
dynamical phenomena may therefore uncover new mechanisms for controlling
many-body quantum states.

In this work, we investigate the resonant dynamics of bound pairs in a
one-dimensional extended Hubbard model under the resonant condition ($U=V\gg
\kappa $). We show that two neighboring doublons become dynamically pinned
and consequently serve as immobile scattering centers. By deriving effective
Hamiltonians within the resonant subspaces, we analytically demonstrate two
strikingly different scattering behaviors. A single electron is completely
reflected by the pinned double-doublon, whereas a singlet bound pair is
perfectly transmitted through it via a resonant intermediate state. As a
consequence, the double-doublon functions as an efficient particle-selective
filter capable of separating bound pairs from individual particles. We
further analyze the collision between a doublon and an electron and predict
a filtering process in which propagating bound pairs pass through two pinned
double-doublons while the single electrons remain trapped between them.
Numerical wave-packet simulations fully confirm the analytical predictions.

Our work uncovers an unconventional transport mechanism arising solely from
interaction-induced resonant dynamics. Besides establishing an exactly
analyzable scattering problem in the extended Hubbard model, our results
provide a simple dynamical protocol for generating, filtering, and
manipulating bound-pair excitations, which may be experimentally accessible
in ultracold-atom realizations of strongly correlated lattice systems.

The remainder of this paper is organized as follows. In Sec. \ref{Model and bound pair}, we introduce the extended Hubbard model and discuss the dynamical properties of doublons and double-doublons under the resonance condition and in the strong-interaction regime. In Sec. \ref{Complete reflection and perfect transmission}, we demonstrate complete reflection and perfect transmission of a single electron and a doublon scattering from a double-doublon barrier by deriving the corresponding effective Hamiltonians in the strong-coupling limit. In Sec. \ref{Doublon-electron collision}, we analyze the collision dynamics between an electron and a doublon, revealing that the scattering between such a composite object and a double-doublon can be decomposed into independent scattering processes involving an electron or a doublon with the double-doublon. Finally, Sec. \ref{Summary} summarizes our conclusions.

\begin{figure*}[tbh]
\centering
\includegraphics[width=0.9\textwidth]{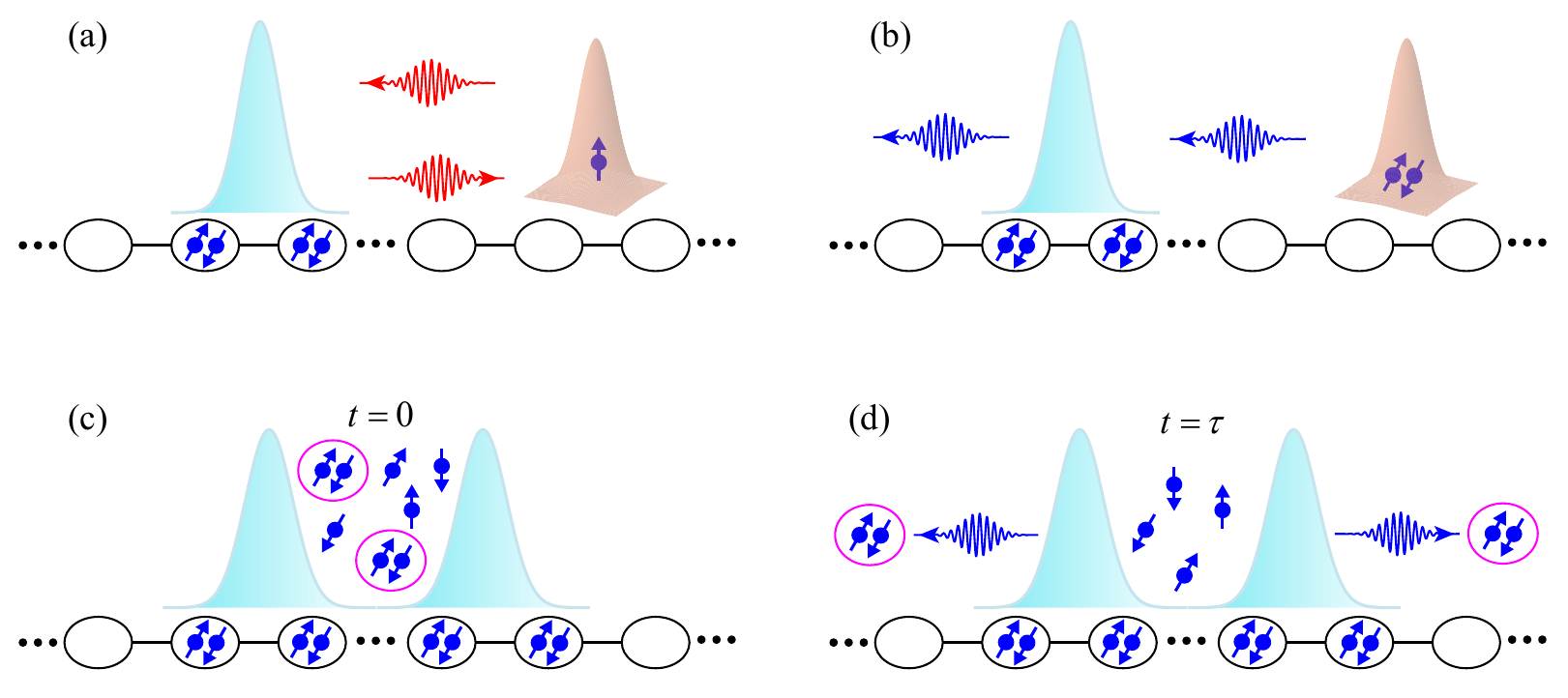}
\caption{Schematic illustration of the main results of this work. The cyan
wave packet represents the barrier, consisting of two neighboring doublon
pairs, while the peach wave packet denotes either a single-electron or a
bound-pair wave packet. (a) Scattering of a single-electron wave packet from
the doublon-pair barrier. We show that the single electron is completely
reflected by the barrier. (b) Scattering of a bound-pair wave packet from
the same barrier. In contrast, the bound pair is perfectly transmitted
through the barrier. (c) and (d) Schematic illustration of a potential
application of these scattering properties. Initially, a mixture of single
electrons and bound pairs is confined within a double-wall structure formed
by two doublon-pair barriers. After sufficiently long time evolution ($%
\protect\tau$), all bound pairs escape from the double-wall system through
perfect transmission, whereas all single electrons remain trapped owing to
complete reflection.}
\label{fig1}
\end{figure*}

\section{Model and bound pair}

\label{Model and bound pair}

We consider an extended Hubbard model with nearest-neighbor interactions on
a one-dimensional chain. The Hamiltonian is given by 
\begin{equation}
H=H_{\mathrm{T}}+H_{\mathrm{V}},
\end{equation}%
where 
\begin{equation}
H_{\mathrm{T}}=\kappa \sum\limits_{j=1}^{N-1}\sum\limits_{\sigma =\uparrow
,\downarrow }c_{j,\sigma }^{\dag }c_{j+1,\sigma }+\mathrm{H.c.},
\end{equation}%
describes nearest-neighbor hopping with amplitude $\kappa $, and 
\begin{equation}
H_{\mathrm{V}}=V\sum\limits_{j=1}^{N-1}n_{j}n_{j+1}+U\sum_{j=1}^{N}n_{j,%
\uparrow }n_{j,\downarrow },
\end{equation}%
contains the nearest-neighbor interaction of strength $V$ and the on-site
interaction of strength $U$. Here $c_{j,\sigma }^{\dagger }$ ($c_{j,\sigma }$%
) creates (annihilates) a fermion with spin $\sigma $ at site $j$, $%
n_{j,\sigma }=c_{j,\sigma }^{\dagger }c_{j,\sigma }$ is the corresponding
number operator, and $n_{j}=n_{j,\uparrow }+n_{j,\downarrow }$ is the total
particle number at site $j$. For $V=0$, the model reduces to the standard
Hubbard model. The nearest-neighbor interaction enriches the phase diagram
of the system and makes the extended Hubbard model a useful description of a
variety of strongly correlated materials, including quasi-one-dimensional
cuprate compounds {\cite{qu2022spin}}. For two-dimensional cases including
square and honeycomb lattices, this model has been investigated and shown to
exhibit enhanced superconductivity induced by $V$ \cite%
{Jiang2022enhancing,Peng2023enhanced,cao2025dominant,Kennedy2025extended}.
In addition, the combination of nearest-neighbor and next-nearest-neighbor
interactions can lead to localization and Hilbert space fragmentation {\cite%
{Frey2022Hilbert}}. Unlike previous studies, which mainly focused on the
phase diagram and superconducting properties of the extended Hubbard model,
we investigate its dynamical properties. In the particle-number basis, $H_{%
\mathrm{V}}$ and $H_{\mathrm{T}}$ correspond to the diagonal and
off-diagonal parts of the Hamiltonian, respectively. Throughout this work,
we consider the strong-interaction regime $U,V\gg \kappa $, where $H_{%
\mathrm{T}}$ can be treated as a perturbation, while $H_{\mathrm{V}}$
defines a set of highly degenerate subspaces. The interplay between the
interaction terms and hopping imposes kinetic constraints on particle
motion, giving rise to a variety of novel dynamical phenomena. In the
following, we illustrate these ideas by analyzing the dynamics of
two-particle states.

We first introduce a set of basis states consisting of doublons, singlets,
and triplets: 
\begin{equation}
\left\vert \psi _{2l}\right\rangle =c_{l,\uparrow }^{\dag }c_{l,\downarrow
}^{\dag }\left\vert \mathrm{v}\right\rangle ,
\end{equation}%
for $l\in \lbrack 1,N]$, 
\begin{eqnarray}
\left\vert \psi _{2l+1}\right\rangle &=&\frac{c_{l,\uparrow }^{\dag
}c_{l+1,\downarrow }^{\dag }-c_{l,\downarrow }^{\dag }c_{l+1,\uparrow
}^{\dag }}{\sqrt{2}}\left\vert \mathrm{v}\right\rangle , \\
\left\vert \phi _{2l+1}\right\rangle &=&\frac{c_{l,\uparrow }^{\dag
}c_{l+1,\downarrow }^{\dag }+c_{l,\downarrow }^{\dag }c_{l+1,\uparrow
}^{\dag }}{\sqrt{2}}\left\vert \mathrm{v}\right\rangle .
\end{eqnarray}%
for $l\in \lbrack 1,N-1]$, where $\left\vert \mathrm{v}\right\rangle $
denotes the vacuum state satisfying $c_{j,\sigma }\left\vert \mathrm{v}%
\right\rangle =0$ for all sites $j$ and spin $\sigma $. Here, even
subscripts denote doublon states, whereas odd subscripts denote
nearest-neighbor singlet states. Acting on the basis states, the Hamiltonian
yields 
\begin{eqnarray}
\left( H-U\right) \left\vert \psi _{2l}\right\rangle &=&\sqrt{2}\kappa
\left( \left\vert \psi _{2l-1}\right\rangle +\left\vert \psi
_{2l+1}\right\rangle \right) ,  \notag \\
\left( H-V\right) \left\vert \psi _{2l+1}\right\rangle &\approx &\sqrt{2}%
\kappa \left( \left\vert \psi _{2l}\right\rangle +\left\vert \psi
_{2l+2}\right\rangle \right) ,  \notag \\
\left( H-V\right) \left\vert \phi _{2l+1}\right\rangle &\approx &0,
\end{eqnarray}%
where the approximate equalities result from projecting onto the low-energy
subspace and neglecting virtual states in which the two particles are no
longer nearest neighbors, whose interaction energies differ by $V$.

Second, throughout this work we consider the resonant condition $V=U$. Under
this condition, the states $\left\{ \left\vert \psi _{2l}\right\rangle
,\left\vert \psi _{2l+1}\right\rangle ,\left\vert \phi _{2l+1}\right\rangle
\right\} $ span a degenerate subspace with interaction energy $U$. Within
this subspace, a doublon and a singlet bound pair propagate as composite
quasiparticles, whereas the triplet state $\left\vert \phi
_{2l+1}\right\rangle $ remains approximately localized. Within the subspace
spanned by $\left\{ \left\vert \psi _{2l}\right\rangle ,\left\vert \psi
_{2l+1}\right\rangle \right\} $, the effective Hamiltonian is equivalent to
a uniform tight-binding model. Similar behavior has also been observed in
Refs \cite%
{Lin2014sudden,Zhang2022conductiong,ma2024superconducting,Zhang2025topological,Zhang2025Bloch,He2025pairing}%
.

Although its dispersion has the same functional form as that of a single
particle, the corresponding bandwidth is renormalized. Similar behavior also
occurs in sectors with more particles.\ In particular, two neighboring
doublon pairs can form a bound state, leading to localization of the
composite excitation. The neighboring doublon-pair state is given by 
\begin{equation}
\left\vert \psi _{\mathrm{dd}}\right\rangle =c_{l,\uparrow }^{\dag
}c_{l,\downarrow }^{\dag }c_{l+1,\uparrow }^{\dag }c_{l+1,\downarrow }^{\dag
}\left\vert \mathrm{v}\right\rangle .
\end{equation}%
Under the resonant condition, $H_{\mathrm{V}}\left\vert \psi _{\mathrm{dd}%
}\right\rangle =6U\left\vert \psi _{\mathrm{dd}}\right\rangle $. Any hopping
process generated by $H_{\mathrm{T}}$ takes the system out of this resonant
subspace and changes the interaction energy by $2U$. In the
strong-interaction regime $U\gg \kappa $, these off-resonant processes are
strongly suppressed. As a result, the neighboring doublon pair is
dynamically pinned and acts as a static scattering barrier for propagating
particles. This mechanism underlies the dynamical filtering effects
discussed below.

\begin{figure}[tbh]
\centering
\includegraphics[width=0.38\textwidth]{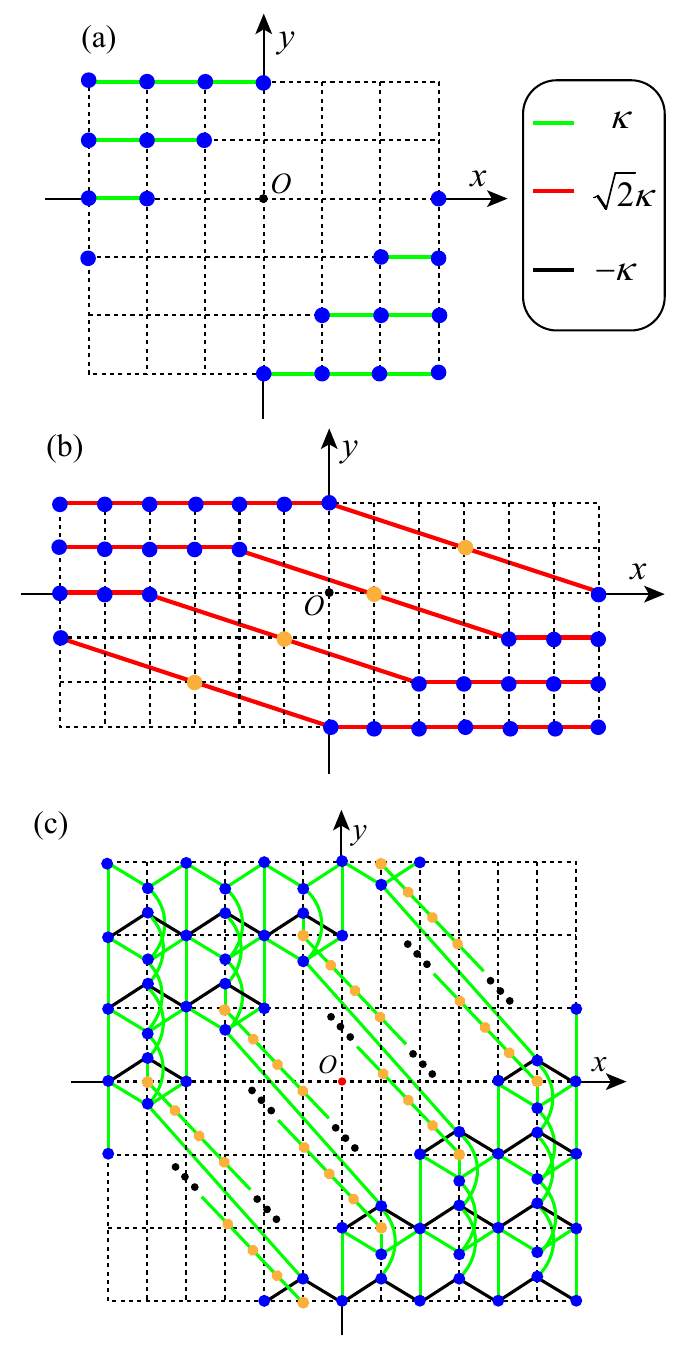}
\caption{Schematics of effective Hamiltonians for several representative
cases under the resonant condition $U=V\gg \protect\kappa $. (a) The
effective Hamiltonian in the subspace constructed by the states including an
electron and a double-doublon, given by Eq. (\protect\ref{Heff_e}). Each
solid blue dot denotes a state defined in Eq. (\protect\ref{stat_e}), where $%
x,y$ give the horizontal and vertical coordinates. The dashed grid
highlights the scale, and the green line marks the hopping strength. (b) The
effective Hamiltonian given in Eq. (\protect\ref{Heff_p}), describing the
interaction between a doublon and a double-doublon. Each solid blue dot
represents a multiple-particle state given by Eq. (\protect\ref{state-p})
mixing a double-doublon and a doublon or a singlet state, and each solid
orange dot represents an intermediate state defined in Eq. (\protect\ref%
{inter_p}). (c) The effective Hamiltonian in the three-particle subspace
with energy $U$, expressed by $H_{\mathrm{eff}}^{\mathrm{e-p}}$ in Eq. (%
\protect\ref{Heff_e_p}). The blue and orange dots are defined in Eq. (%
\protect\ref{group1}) and Eq. (\protect\ref{group2}), respectively. The
chain along the orange dots is an infinite chain expressed by Eq. (\protect
\ref{additional_chain}).}
\label{fig2}
\end{figure}

\section{Complete reflection and perfect transmission}

\label{Complete reflection and perfect transmission}

In the previous section, we showed that a neighboring doublon pair remains
dynamically pinned and thus acts as a static scattering barrier. We next
investigate how single electrons and singlet bound pairs scatter from this
barrier, revealing its particle-filtering capability. The main results of
this section are schematically illustrated in Fig. \ref{fig1}(a) and (b).

\subsection{Perfect electron reflection}

To proceed, we introduce a double-doublon creation operator composed of four
fermionic creation operators, 
\begin{equation}
d_{y}^{\dag }=c_{y,\uparrow }^{\dag }c_{y,\downarrow }^{\dag
}c_{y+1,\uparrow }^{\dag }c_{y+1,\downarrow }^{\dag },
\end{equation}%
representing two neighboring doublons centered around position $y$. The
corresponding basis states are constructed as 
\begin{equation}
\left\vert x,y\right\rangle _{\mathrm{e}}^{\mathrm{dd}}=c_{x,\sigma }^{\dag
}d_{y}^{\dag }\left\vert \mathrm{v}\right\rangle ,  \label{stat_e}
\end{equation}%
which describe a composite system consisting of a double-doublon and an
additional electron. Here, $x$ and $\sigma $ denote the coordinate and spin
of the electron, respectively. We focus on the $6U$-energy subspace, in
which the double-doublon and the electron cannot overlap due to the
interaction constraint. Consequently, the allowed configurations are
restricted to $x\geq y+3$ or $x\leq y-2$. The projected Hamiltonian in this
subspace describes the hopping motion of the electron around a fixed
double-doublon: 
\begin{eqnarray}
&&H_{\mathrm{eff}}^{\mathrm{dd-e}}=\sum\limits_{y}\kappa \lbrack \sum_{x\geq
y+3}\left\vert x,y\right\rangle _{\mathrm{e}}^{\mathrm{dd}}\left\langle
x+1,y\right\vert _{\mathrm{e}}^{\mathrm{dd}}  \notag \\
&&+\sum_{x\leq y-2}\left\vert x-1,y\right\rangle _{\mathrm{e}}^{\mathrm{dd}%
}\left\langle x,y\right\vert _{\mathrm{e}}^{\mathrm{dd}}]+\mathrm{H.c.}.
\label{Heff_e}
\end{eqnarray}%
This Hamiltonian provides a clear physical picture: for a given
double-doublon located at position y, the effective chain is divided into
two disconnected sectors, each corresponding to a uniform tight-binding
chain with a boundary at $y-2$ or $y+3$. In Fig. \ref{fig2}(a), we
illustrate the connectivity structure of $H_{\mathrm{eff}}^{\mathrm{dd-e}}$
for a finite system by mapping each basis state $\left\vert x,y\right\rangle
_{\mathrm{e}}^{\mathrm{dd}}$ onto a dot with coordinates $(x,y)$. This
demonstrates that, in the strong-interaction limit $U\gg \kappa $, the
double-doublon acts as an impenetrable barrier, resulting in perfect
reflection of the electron.

\subsection{Perfect doublon transmission}

We now investigate the scattering of a doublon, viewed as a composite bound
pair, from a double-doublon barrier. The analysis in Sec. \ref{Model and
bound pair} shows that the two electrons forming a doublon propagate
coherently along the chain. A natural question then arises: what happens
when a doublon collides with a double-doublon? In the following, we address
this question by analyzing the corresponding effective Hamiltonian.

First, we introduce a compact representation of the singlet and doublon
states, 
\begin{eqnarray}
\left\vert 2l-1\right\rangle _{\mathrm{p}} &=&\frac{c_{l-1,\uparrow }^{\dag
}c_{l,\downarrow }^{\dag }-c_{l-1,\downarrow }^{\dag }c_{l,\uparrow }^{\dag }%
}{\sqrt{2}}\left\vert \mathrm{v}\right\rangle,  \notag \\
\left\vert 2l\right\rangle _{\mathrm{p}} &=&c_{l,\uparrow }^{\dag
}c_{l,\downarrow }^{\dag }\left\vert \mathrm{v}\right\rangle ,  \label{pair}
\end{eqnarray}%
which span all accessible states during the propagation of a doublon. We
then construct the composite six-particle states consisting of a
double-doublon and a pair state $\left\vert x\right\rangle _{\mathrm{p}}$, 
\begin{equation}
\left\vert x,y\right\rangle _{\mathrm{p}}^{\mathrm{dd}}=d_{y}^{\dag
}\left\vert x\right\rangle _{\mathrm{p}},  \label{state-p}
\end{equation}%
with $x\geq 2y+6$ or $x\leq 2y-4$. These constraints ensure that the
double-doublon and the propagating pair do not overlap, such that all these
states are degenerate with energy $7U$ under $H_{\mathrm{V}}$. There exist
additional states in this energy subspace that cannot be represented by the
above construction. We define such an intermediate state as 
\begin{equation}
\left\vert 2y+1,y\right\rangle _{\mathrm{p}}^{\text{\textrm{dd}}%
}=c_{y-1,\uparrow }^{\dag }c_{y-1,\downarrow }^{\dag }c_{y+2,\uparrow
}^{\dag }c_{y+2,\downarrow }^{\dag }\left\vert 2y+1\right\rangle _{\mathrm{p}%
},  \label{inter_p}
\end{equation}%
where 
\begin{equation}
\left\vert 2y+1\right\rangle _{\mathrm{p}}=\frac{c_{y,\uparrow
}^{\dag}c_{y+1,\downarrow }^{\dag }-c_{y,\downarrow }^{\dag }c_{y+1,\uparrow
}^{\dag }}{\sqrt{2}}\left\vert \mathrm{v}\right\rangle
\end{equation}%
is defined according to Eq. (\ref{pair}). Physically, this state represents
an intermediate configuration arising when the doublon $c_{y+2,\uparrow
}^{\dag }c_{y+2,\downarrow }^{\dag }$ encounters the double-doublon $%
c_{y-1,\uparrow }^{\dag }c_{y-1,\downarrow }^{\dag }c_{y,\uparrow }^{\dag
}c_{y,\downarrow }^{\dag }$. The hopping Hamiltonian $H_{\mathrm{T}}$
couples this intermediate state to the above six-particle states as%
\begin{eqnarray}
H_{\mathrm{T}}\left\vert 2y+1,y\right\rangle _{\mathrm{p}}^{\text{\textrm{dd}%
}} &=&\sqrt{2}\kappa \left\vert 2y+4,y-1\right\rangle _{\mathrm{p}}^{\mathrm{%
dd}}  \notag \\
&&+\sqrt{2}\kappa \left\vert 2y-2,y+1\right\rangle _{\mathrm{p}}^{\mathrm{dd}%
}.
\end{eqnarray}%
Therefore, the doublon can pass through the double-doublon barrier, leading
to an exchange of their spatial positions. In contrast to the
single-electron case, where the double-doublon acts as an impenetrable
barrier, a bound doublon can resonantly pass through it via the intermediate
state.

Indeed, a direct derivation yields the effective Hamiltonian 
\begin{eqnarray}
H_{\mathrm{eff}}^{\mathrm{dd-p}} &=&\sqrt{2}\kappa
\sum\limits_{y}(\sum\limits_{x\geq 2y+6}\left\vert x,y\right\rangle _{%
\mathrm{p}}\left\langle x+1,y\right\vert _{\mathrm{p}}  \notag \\
&&+\sum\limits_{x\leq 2y}\left\vert x-1,y+2\right\rangle _{\mathrm{p}%
}\left\langle x,y+2\right\vert _{\mathrm{p}}  \notag \\
&&+\left\vert 2y+3,y+1\right\rangle _{\mathrm{p}}\left\langle
2y+6,y\right\vert _{\mathrm{p}}  \notag \\
&&+\left\vert 2y,y+2\right\rangle _{\mathrm{p}}\left\langle
2y+3,y+1\right\vert _{\mathrm{p}}+\mathrm{H.c.}).  \label{Heff_p}
\end{eqnarray}%
Similar to the single-electron case, we illustrate the schematic
representation of $H_{\mathrm{eff}}^{\mathrm{dd-p}}$ for a finite system in
Fig. \ref{fig2}(b). The effective Hamiltonian consists of a series of
uniform tight-binding chains labeled by the position of the double-doublon.
Each chain describes a scattering process in which a doublon colliding with
a double-doublon is perfectly transmitted through the barrier, accompanied
by a displacement of the double-doublon by two lattice sites. This behavior
is qualitatively different from the single-electron case, where the
double-doublon acts as an impenetrable barrier and the electron undergoes
perfect reflection.

\begin{figure*}[tbh]
\centering
\includegraphics[width=1.0\textwidth]{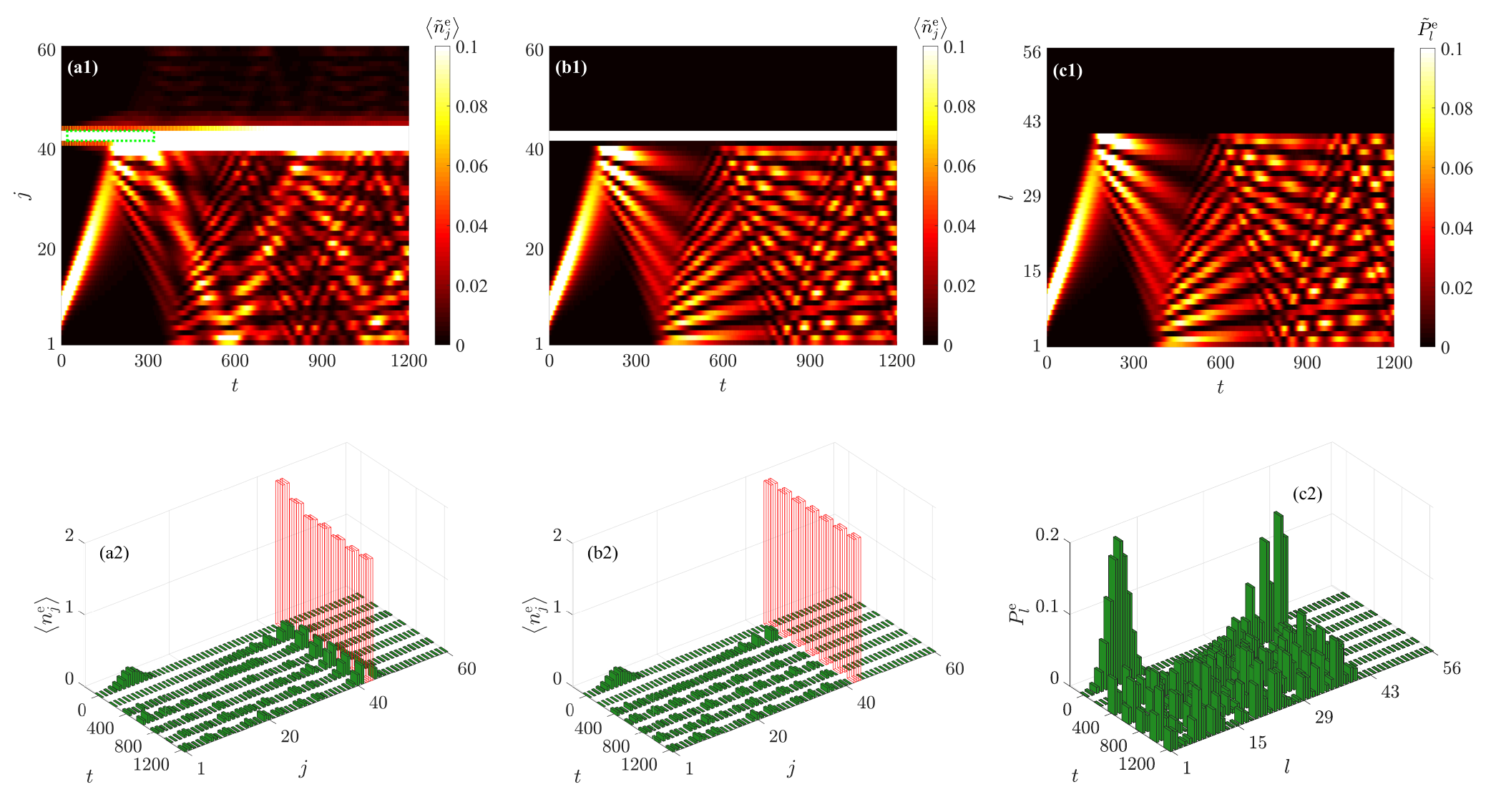}
\caption{Numerical results for single-electron wave packet dynamics. The
initial state is given by Eq. (\protect\ref{initial_e}) with $j_{0}=8$, $%
y_{0}=42$, $k_{0}=\protect\pi /4$, and $\protect\alpha=4$. The hopping
strength in $H_{\mathrm{T} }$ is set to $\protect\kappa =0.1$, and $N=60$.
Panels (a1) and (b1) show the mean particle number at each site defined in
Eq. (\protect\ref{n_e}) for $U=0.5$ and $U=5$, respectively. Here $%
\left\langle \tilde{n}_{j}^{\mathrm{e} }\right\rangle $ is the reduced mean
particle number, obtained by dividing $\left\langle n_{j}^{\mathrm{e}%
}\right\rangle $ by a constant. The green dashed rectangle in panel (a1)
marks the position of the double-doublon. Panel (c1) shows the dynamics
performed with the effective Hamiltonian in Eq. (\protect\ref{Heff_e}) with
the same initial state as in (a1) and (b1), and $\left\langle \tilde{P}_{l}^{%
\mathrm{e}}\right\rangle $ is the reduced probability, proportional to $%
\left\langle {P}_{l}^{\mathrm{e}}\right\rangle $ in Eq. (\protect\ref{P_e}).
Panels (a2), (b2), and (c2) demonstrate the corresponding details of panels
(a1), (b1), and (c1) for several specific time points; the red-edged bar
marks the double-doublon.}
\label{fig3}
\end{figure*}

\section{Doublon-electron collision}

\label{Doublon-electron collision}

In the previous section, we demonstrated that the double-doublon barrier
exhibits distinct responses to different incident particles: a doublon is
perfectly transmitted, while an electron is perfectly reflected. This
suggests that the double-doublon can function as a particle-selective
filter. We now turn to the complementary problem of electron-doublon
scattering.

Following the above analysis, we define a set of three-particle states in
the energy-$U$ subspace, 
\begin{eqnarray}
\overline{\left\vert x,y\right\rangle }_{1} &=&\left\vert x,y\right\rangle
=c_{y,\uparrow }^{\dag }c_{x/2,\uparrow }^{\dag }c_{x/2,\downarrow }^{\dag
}\left\vert \mathrm{v}\right\rangle ,  \notag \\
\overline{\left\vert x,y\right\rangle }_{2} &=&\left\vert x-1,y+\frac{1}{3}%
\right\rangle =c_{y,\uparrow }^{\dag }c_{x/2-1,\downarrow }^{\dag
}c_{x/2,\uparrow }^{\dag }\left\vert \mathrm{v}\right\rangle ,  \notag \\
\overline{\left\vert x,y\right\rangle }_{3} &=&\left\vert x-1,y-\frac{1}{3}%
\right\rangle =c_{y,\uparrow }^{\dag }c_{x/2-1,\uparrow }^{\dag
}c_{x/2,\downarrow }^{\dag }\left\vert \mathrm{v}\right\rangle ,
\label{group1}
\end{eqnarray}%
which is a set of basis states for the electron-doublon scattering subspace,
with $x\geq 2y+4$ or $x\leq 2y-4$ and $x\in \mathrm{even}$. Each state is
labeled by a pair of coordinates $(a,b)$ in the $x$-$y$ plane. The state $%
\overline{\left\vert x,y\right\rangle }_{1}$ describes a composite
configuration consisting of a spin-up electron and a doublon. For
simplicity, we fix the spin of the electron to be $\uparrow $. Such a
configuration can hop to $\overline{\left\vert x,y\right\rangle }_{2}$ and $%
\overline{\left\vert x,y\right\rangle }_{3}$ while remaining within the same
energy sector.

Acting with $H_{\mathrm{T}}$ on these states generates another set of basis
states within the same energy subspace: 
\begin{eqnarray}
\left\vert \psi _{1}\left( l,y\right) \right\rangle &=&c_{y-3,\uparrow
}^{\dag }c_{y-2,\uparrow }^{\dag }c_{l,\downarrow }^{\dag }\left\vert 
\mathrm{v}\right\rangle ,\quad l\geq y,  \notag \\
\left\vert \psi _{2}\left( l,y\right) \right\rangle &=&c_{l-3,\downarrow
}^{\dag }c_{y-1,\uparrow }^{\dag }c_{y,\uparrow }^{\dag }\left\vert \mathrm{v%
}\right\rangle ,\quad l\leq y.  \label{group2}
\end{eqnarray}%
The states $\left\vert \psi _{1}\left( l,y\right) \right\rangle $ and $%
\left\vert \psi _{2}\left( l,y\right) \right\rangle $ span two semi-infinite
chains, along which the electron created by $c_{l,\downarrow }^{\dag }$ and $%
c_{l-3,\downarrow }^{\dag }$ can propagate. During this process, the
two-electron configurations $c_{y-3,\uparrow }^{\dag }c_{y-2,\uparrow
}^{\dag }$ and $c_{y-1,\uparrow }^{\dag }c_{y,\uparrow }^{\dag }$ remain
localized. In particular, for the special case $l=y$, we assign coordinates
to the corresponding boundary states as 
\begin{eqnarray}
\left\vert \psi _{1}\left( y,y\right) \right\rangle &=&c_{y-3,\uparrow
}^{\dag }c_{y-2,\uparrow }^{\dag }c_{y,\downarrow }^{\dag }\left\vert 
\mathrm{v}\right\rangle =\left\vert 2y-1,y-3\right\rangle ,  \notag \\
\left\vert \psi _{2}\left( y,y\right) \right\rangle &=&c_{y-3,\downarrow
}^{\dag }c_{y-1,\uparrow }^{\dag }c_{y,\uparrow }^{\dag }\left\vert \mathrm{v%
}\right\rangle =\left\vert 2y-5,y\right\rangle .
\end{eqnarray}%
Combining these two sets of states, the effective Hamiltonian takes the form 
\begin{equation}
H_{\mathrm{eff}}^{\mathrm{e-p}}=\kappa \left( H_{1}+H_{2}+H_{3}\right) .
\label{Heff_e_p}
\end{equation}%
Here 
\begin{eqnarray}
H_{1} &=&\sum_{y}[\sum\limits_{x\leq 2y-4}^{x\in \mathrm{even}}(\overline{%
\left\vert x,y\right\rangle }_{1}\overline{\left\langle x,y\right\vert }_{3}-%
\overline{\left\vert x,y\right\rangle }_{1}\overline{\left\langle
x,y\right\vert }_{2}  \notag \\
&&+\overline{\left\vert x-2,y\right\rangle }_{1}\overline{\left\langle
x,y\right\vert }_{3}-\overline{\left\vert x-2,y\right\rangle }_{1}\overline{%
\left\langle x,y\right\vert }_{2}  \notag \\
&&+\sum\limits_{j=1}^{3}\overline{\left\vert x,y+1\right\rangle }_{j}%
\overline{\left\langle x,y\right\vert }_{j})+\sum_{x\geq 2y+4}^{x\in \mathrm{%
even}}(\overline{\left\vert x,y\right\rangle }_{1}\overline{\left\langle
x+2,y\right\vert }_{3}  \notag \\
&&-\overline{\left\vert x,y\right\rangle }_{1}\overline{\left\langle
x+2,y\right\vert }_{2}+\overline{\left\vert x+2,y\right\rangle }_{1}%
\overline{\left\langle x+2,y\right\vert }_{3}  \notag \\
&&-\overline{\left\vert x+2,y\right\rangle }_{1}\overline{\left\langle
x+2,y\right\vert }_{2}+\sum\limits_{j=1}^{3}\overline{\left\vert
x,y-1\right\rangle }_{j}\overline{\left\langle x,y\right\vert }_{j})  \notag
\\
&&+\overline{\left\vert 2y,y-3\right\rangle }_{2}\overline{\left\langle
2y-4,y\right\vert }_{3}]+\mathrm{H.c.},
\end{eqnarray}%
describes the hopping processes within the first set of states introduced in
Eq. (\ref{group1}). In particular, the term $\overline{\left\vert
2y,y-3\right\rangle }_{2}\overline{\left\langle 2y-4,y\right\vert }_{3}$
represents the process in which the electron is transmitted through the
doublon. The second contribution is given by 
\begin{eqnarray}
H_{2} &=&\sum_{y}(\left\vert \psi _{1}\left( y,y\right) \right\rangle 
\overline{\left\langle 2y,y-3\right\vert }_{3}  \notag \\
&&+\left\vert \psi _{2}\left( y,y\right) \right\rangle \overline{%
\left\langle 2y-4,y\right\vert }_{2})+\mathrm{H.c.},
\end{eqnarray}%
which describes the intermediate processes connecting the first set of
states in Eq. (\ref{group1}) and the second set of states in Eq. (\ref%
{group2}). Finally, 
\begin{eqnarray}
&&H_{3}=\sum_{l\geq y}\left\vert \psi _{1}\left( l,y\right) \right\rangle
\left\langle \psi _{1}\left( l+1,y\right) \right\vert  \notag \\
&&+\sum_{l\leq y}(\left\vert \psi _{2}\left( l-1,y\right) \right\rangle
\left\langle \psi _{2}\left( l,y\right) \right\vert )+\mathrm{H.c.},
\label{additional_chain}
\end{eqnarray}%
describes the hopping processes within the second set of states introduced
in Eq. (\ref{group2}). The term $H_{3}$ describes the transmission channel
of the electron-doublon scattering process, where the spin-down electron
propagates through the doublon formed by two spin-up electrons.

To clarify the physical picture, Fig. \ref{fig2}(c) presents a schematic
representation of the effective Hamiltonian $H_{\mathrm{eff}}^{\mathrm{e-p}}$
in the $x$-$y$ plane. It shows that three scattering channels coexist when a
spin-$\uparrow $ electron collides with a doublon: (i) the electron is
reflected by the doublon while leaving behind a mobile pair of electrons
with opposite spins; (ii) the electron is transmitted through the doublon,
again leaving a mobile pair of electrons with opposite spins; and (iii) a
spin-$\downarrow $ electron is transmitted through the doublon, leaving
behind an immobile pair of electrons with the same spin. The mobile pair
with opposite spins may subsequently recombine into a doublon. These results
show that, when the three-particle state consisting of an electron and a
doublon is well separated from the double-doublon, its scattering dynamics
can be decomposed into independent scattering processes of either the
electron or the doublon with the double-doublon, as discussed in Sec. \ref%
{Complete reflection and perfect transmission}.

\begin{figure*}[tbh]
\centering
\includegraphics[width=1.0\textwidth]{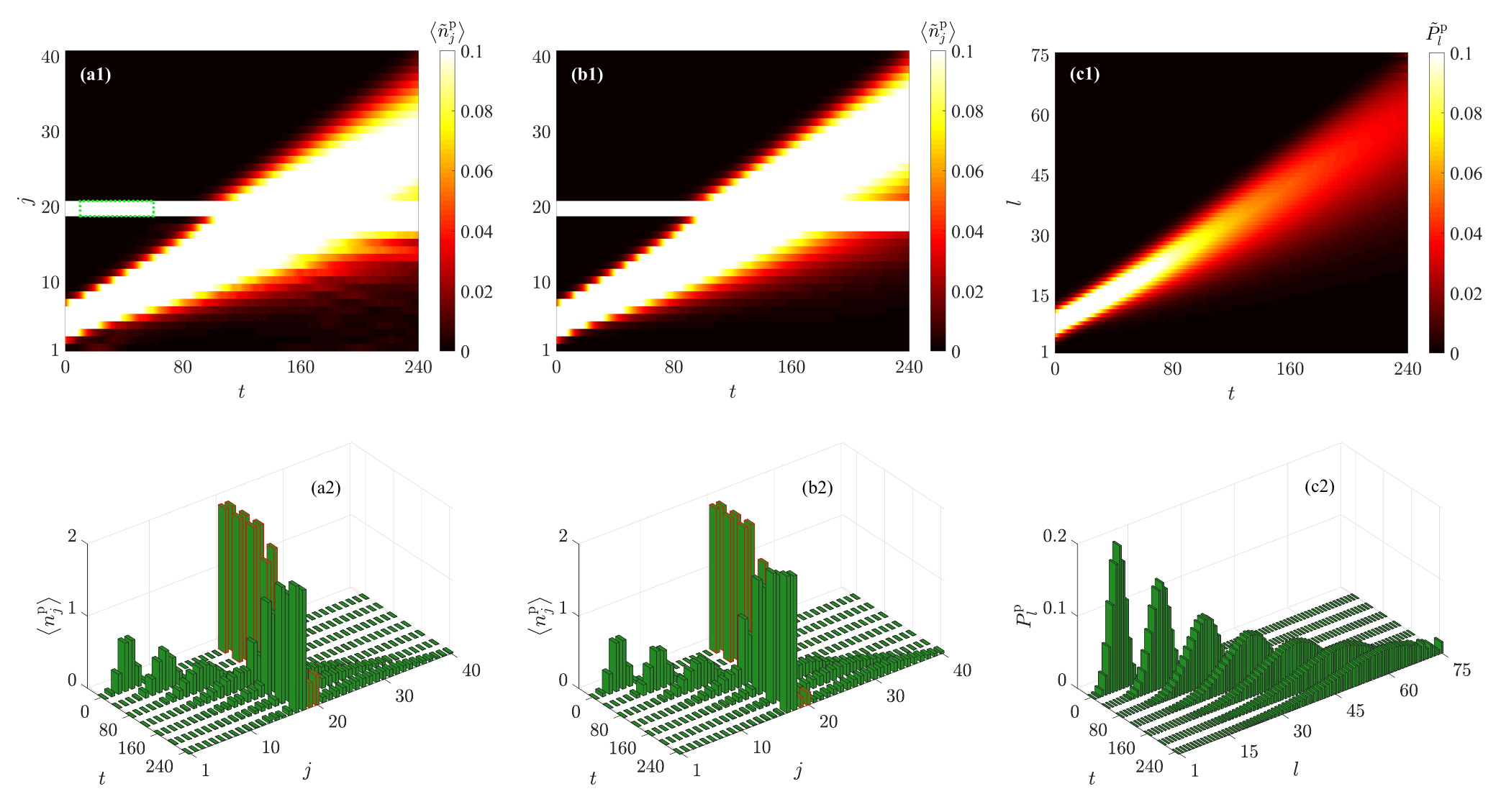}
\caption{Numerical results for two-electron wave packet dynamics. The
initial state is given by Eq. (\protect\ref{initial_p}) with $j_{0}=8$, $%
y_{0}=19$, $k_{0}=\protect\pi /4$, and $\protect\alpha=4$. The hopping
strength in $H_{\mathrm{T}} $ is set to $\protect\kappa=0.1$, and $N=40$.
Panels (a1) and (b1) show the mean particle number at each site defined in
Eq. (\protect\ref{n_p}) for $U=1 $ and $U=5$, respectively. $\left\langle 
\tilde{n}_{j}^{\mathrm{p}}\right\rangle$ is the reduced mean particle
number, similar to that in Fig. \protect\ref{fig3}. The green dashed
rectangle in panel (a1) marks the position of the double-doublon. Panel (c1)
shows the dynamics performed with the effective Hamiltonian in Eq. (\protect
\ref{Heff_p}) with the same initial state as in (a1) and (b1), and $%
\left\langle \tilde{P}_{l}^{\mathrm{p}}\right\rangle$ is the reduced
probability, proportional to $\left\langle {P}_{l}^{\mathrm{p}}\right\rangle$
in Eq. (\protect\ref{P_p}). Panels (a2), (b2), and (c2) demonstrate the
corresponding details of panels (a1), (b1), and (c1) for several specific
time points; the red-edged bar marks the initial position of the
double-doublon.}
\label{fig4}
\end{figure*}

\section{Numerical results}

\label{Numerical results}

In the previous sections, we analyzed the collision between a bound pair of
doublons and an electron or a doublon, showing that total reflection and
perfect transmission occur in the resonant regime $U=V\gg \kappa $. Such
scattering processes can be realized by wave packets carrying a well-defined
momentum. To confirm these analytical predictions, we perform wave-packet
simulations. We first consider the single-electron wave packet illustrated
in Fig. \ref{fig1}(a) as the initial state,
\begin{equation}
\left\vert \psi _{\mathrm{e}}\left( 0\right) \right\rangle =\Omega _{\mathrm{%
e}}^{-1}\sum_{j=1}^{15}e^{-ik_{0}j}e^{-(j-j_{0})^{2}/\alpha
^{2}}c_{j,\uparrow }^{\dag }d_{y_{0}}^{\dag }\left\vert \mathrm{v}%
\right\rangle ,  \label{initial_e}
\end{equation}%
where $\Omega _{\mathrm{e}}^{-1}$ is the normalization constant. The initial
state belongs to the subspace with energy $6U$. Its time evolution is given
by 
\begin{equation}
\left\vert \psi _{\mathrm{e}}\left( t\right) \right\rangle
=e^{-iHt}\left\vert \psi _{\mathrm{e}}\left( 0\right) \right\rangle .
\end{equation}%
To simulate larger systems while maintaining computational efficiency, we
employ an energy-based Hilbert-space truncation. Starting from the basis
states contained in the effective Hamiltonian in Eq. (\ref{Heff_e}), we
iteratively apply $H_{\mathrm{T}}$ to generate additional basis states.
States whose energies fall within the window $[4U,8U]$ are retained, whereas
all others are discarded. The procedure is continued until the size of the
truncated Hilbert space reaches a prescribed cutoff. The Hamiltonian is then
constructed in this truncated basis and used to calculate the real-time
evolution.

We define the local particle density 
\begin{equation}
\left\langle n_{j}^{\mathrm{e}}\right\rangle =\left\langle \psi _{\mathrm{e}%
}\left( t\right) \right\vert n_{j}\left\vert \psi _{\mathrm{e}}\left(
t\right) \right\rangle ,  \label{n_e}
\end{equation}%
to characterize the propagation of the wave packet. For comparison with the
analytical prediction, we also simulate the time evolution governed by the
effective Hamiltonian in Eq. (\ref{Heff_e}) using the same initial state.
The corresponding evolved state is 
\begin{equation}
\left\vert \Psi _{\mathrm{e}}\left( t\right) \right\rangle =e^{-itH_{\mathrm{%
eff}}^{\mathrm{dd-e}}}\left\vert \psi _{\mathrm{e}}\left( 0\right)
\right\rangle .
\end{equation}%
We then calculate the probability distribution 
\begin{equation}
P_{l}^{\mathrm{e}}=\left\vert \left\langle \Psi _{\mathrm{e}}\left( t\right)
|\psi _{l}^{\mathrm{e}}\right\rangle \right\vert ^{2},  \label{P_e}
\end{equation}%
where $\left\vert \psi _{l}^{\mathrm{e}}\right\rangle $ denotes the basis
state associated with the $l$th site of the effective chain. The numerical
results are presented in Fig. \ref{fig3}. For small $U$, the electron
repeatedly reflects from the system boundaries and the double doublon, while
exhibiting a finite probability of transmitting through the latter. As $U$
increases, the transmission probability is progressively suppressed. In the
large-$U$ limit, the electron undergoes complete reflection from the double
doublon, in excellent agreement with the prediction of the effective
Hamiltonian.

Similar to the single-electron case, we consider a two-particle wave packet
to simulate the interaction between a double-doublon and a doublon, as
illustrated in Fig. \ref{fig1}(b). The initial state is given by
\begin{equation}
\left\vert \psi _{\mathrm{p}}\left( 0\right) \right\rangle =\Omega _{\mathrm{%
p}}^{-1}\sum_{j=1}^{15}e^{-ik_{0}j}e^{-(j-j_{0})^{2}/\alpha
^{2}}d_{y_{0}}^{\dag }\left\vert j\right\rangle _{\mathrm{p}},
\label{initial_p}
\end{equation}%
where $\left\vert j\right\rangle _{\mathrm{p}}$ denotes the two-particle
state defined in Eq. (\ref{pair}). Following the same truncation procedure
described above, we retain the states whose energies lie within the interval 
$[6U,8U]$ to construct the truncated Hamiltonian. The evolved state is then
obtained as 
\begin{equation}
\left\vert \psi _{\mathrm{p}}\left( t\right) \right\rangle
=e^{-iHt}\left\vert \psi _{\mathrm{p}}\left( 0\right) \right\rangle ,
\end{equation}%
and the corresponding local particle density is%
\begin{equation}
\left\langle n_{j}^{\mathrm{p}}\right\rangle =\left\langle \psi _{\mathrm{p}%
}\left( t\right) \right\vert n_{j}\left\vert \psi _{\mathrm{p}}\left(
t\right) \right\rangle .  \label{n_p}
\end{equation}%
Similarly, we simulate the time evolution governed by the effective
Hamiltonian in Eq. (\ref{Heff_p}) using the same initial state. The
corresponding evolved state is 
\begin{equation}
\left\vert \Psi _{\mathrm{p}}\left( t\right) \right\rangle =e^{-itH_{\mathrm{%
eff}}^{\mathrm{dd-p}}}\left\vert \psi _{\mathrm{p}}\left( 0\right)
\right\rangle ,
\end{equation}%
and the probability distribution is calculated as%
\begin{equation}
P_{l}^{\mathrm{p}}=\left\vert \left\langle \Psi _{\mathrm{p}}\left( t\right)
|\psi _{l}^{\mathrm{p}}\right\rangle \right\vert ^{2},  \label{P_p}
\end{equation}%
where $\left\vert \psi _{l}^{\mathrm{p}}\right\rangle $ denotes the basis
state associated with the $l$th site of the effective chain. The numerical
results are presented in Fig. \ref{fig4}. In the large-$U$ limit, the
two-particle wave packet is transmitted through the double doublon with
nearly unit probability, while the double doublon is displaced by two
lattice sites, in excellent agreement with the analytical prediction. Even
for relatively small values of $U$, only minor deviations from the ideal
behavior are observed.

Based on the analysis in Sec. \ref{Doublon-electron collision}, we predict
another intriguing dynamical process. A three-particle state composed of a
doublon and an electron is initially prepared between two well-separated
double doublons [Fig. \ref{fig1}(c)]. At long times, the two-particle wave
packet is expected to pass through both double doublons, while two single
electrons remain confined in the double-wall structure [Fig. \ref{fig1}(d)].
Owing to the prohibitively large Hilbert-space dimension required for a
direct simulation, we restrict ourselves here to presenting the underlying
physical picture, leaving a quantitative numerical verification for future
work.

\section{Summary}

\label{Summary}

We have investigated the resonant dynamics of bound pairs in a
one-dimensional extended Hubbard model in the strong-interaction regime.
Under the resonance condition ($U=V\gg \kappa $), we demonstrated that two
neighboring doublons become dynamically pinned and act as an effective
scattering center for propagating quasiparticles. By deriving effective
Hamiltonians in different resonant subspaces, we obtained an analytical
description of several scattering processes involving electrons, doublons,
and double-doublons.

Our analysis reveals a remarkable particle-selective scattering mechanism.
While a single electron is completely reflected by the pinned
double-doublon, a singlet bound pair undergoes perfect transmission through
a resonant intermediate state. This qualitative difference originates from
the distinct connectivity of the corresponding effective Hamiltonians rather
than from single-particle interference. Building upon these results, we
further proposed a three-particle dynamical process in which an electron and
a doublon are spatially separated by two pinned double-doublons, leading to
the transmission of bound pairs and the confinement of single particles.
Numerical wave-packet simulations are in excellent agreement with the
analytical predictions and demonstrate that the ideal scattering behavior is
rapidly approached as the interaction strength increases.

These results establish the neighboring double-doublon as an efficient
dynamical filter that distinguishes composite bound states from individual
particles solely through interaction-induced resonant dynamics. Our work
therefore uncovers an unconventional transport mechanism in strongly
correlated systems and provides a simple route toward the coherent
generation, separation, and manipulation of bound-pair excitations. The
underlying mechanism is expected to be applicable to a broader class of
constrained quantum many-body systems and may be explored experimentally in
ultracold-atom realizations of extended Hubbard models.

\acknowledgments This work was supported by the National Natural Science
Foundation of China (Grant No. 12374461).

\bibliography{PairfilterReference}

\end{document}